\documentstyle[epsfig]{aipproc}
\begin{document}
\title{ZeV Air Showers: \\
               The View from Auger} 

\author{Enrique Zas}
\address{Departamento de F\'\i sica de Part\'\i culas,\\
Universidade de Santiago de Compostela, E-15706 Santiago, Spain.\\
zas@fpaxp1.usc.es}

\maketitle

\begin{abstract}
In this article I briefly discuss the characteristics of the Auger 
observatories paying particular attention to the role of inclined 
showers, both in the search for high energy neutrino interactions deep 
in the atmosphere and as an alternative tool for the study of cosmic rays, 
particulartly their composition. 

\end{abstract}


%
%
%
%

\section{Introduction}

The detection of high energy showers with the radio technique has 
been shown to have a high potential for astroparticle physics thoroughout 
this conference. 
One of the advantages of radio detection is that provided an appropriate wavelength can be chosen, exceeding the shower dimensions, 
the emission from all the shower particles becomes coherent. 
When the emission from all particles is coherent the emitted power should 
scale with the square of the primary energy. The technique thus becomes 
most advantageous for the detection of the highest energy particles. 
The detection of air showers with the radio technique was started in the 
1950's \cite{weekes} but it experienced many difficulties and 
other methods took the leading role in cosmic ray detection, at first arrays 
of particle detectors and \v Cerenkov telescopes and more recently air 
fluorescence detectors.  

\begin{figure}[hb] 
\center{
\centerline{\epsfig{file=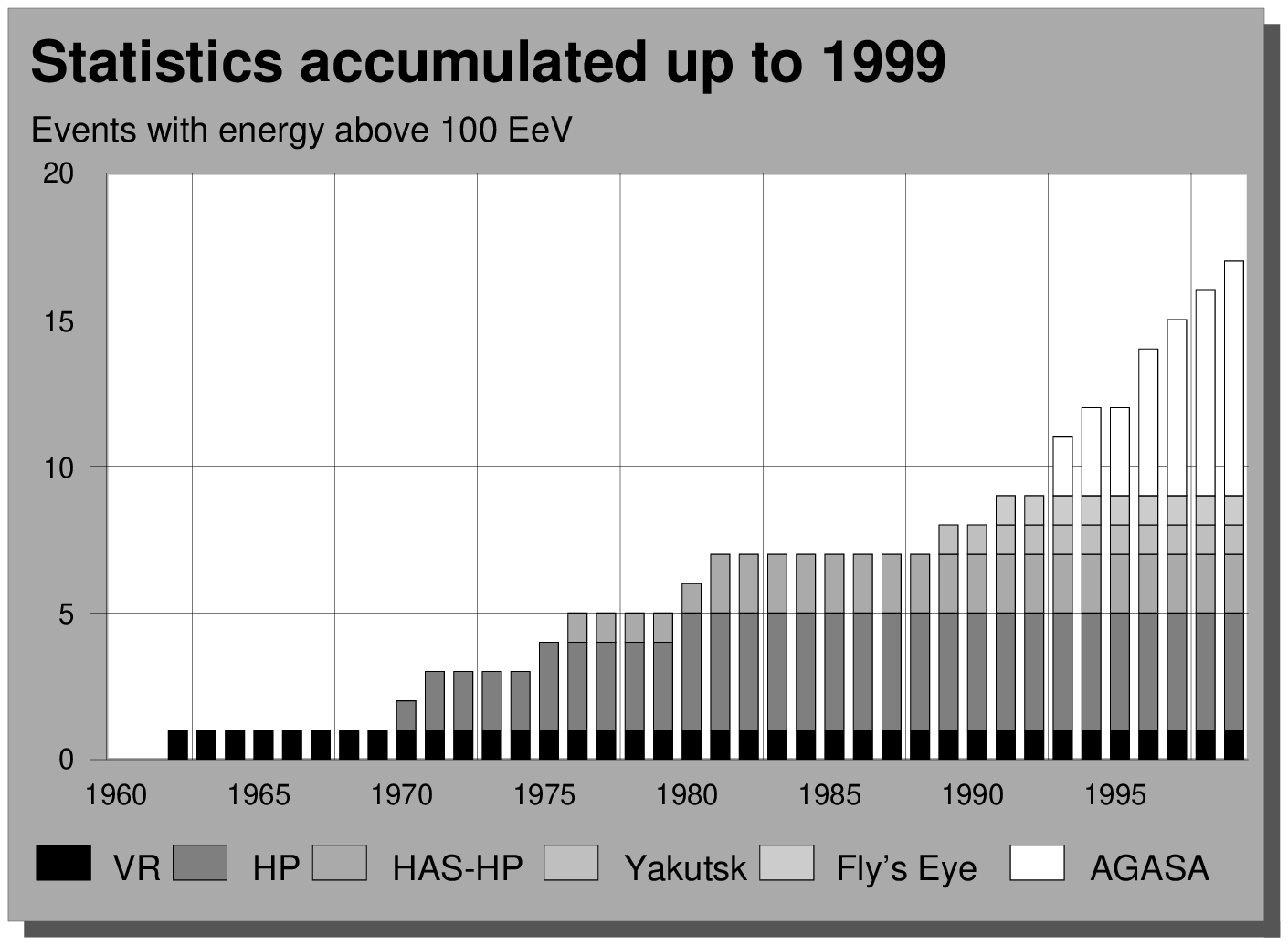,height=2.in,width=2.9in} 
\epsfig{file=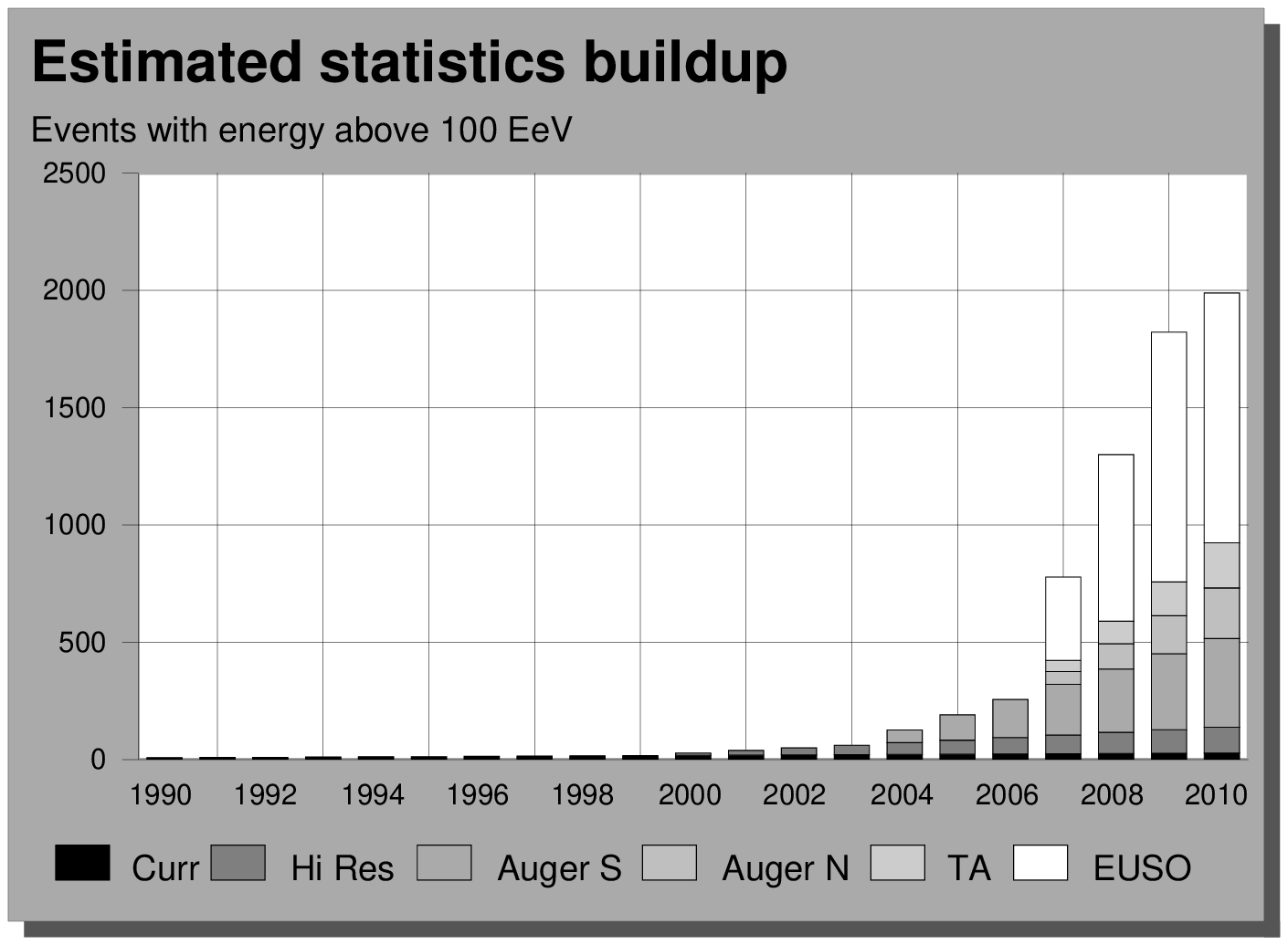,height=2.in,width=2.9in}}  }
\vspace{10pt}
\caption{Left: Events with $E>10^{20}$~eV 
detected by different experiments: 
Volcano Ranch (VR), Haverah Park (HP), Horizontal Air Showers in 
Haverah Park (HAS-HP), Yakutsk, Fly's Eye and AGASA. Right:  
Data buildup after 2000 from: HiRes, 
South and North Auger observatories (Auger S,N), Telescope 
Array (TA) and EUSO.}
\label{uptonow}
\end{figure}

One of the most intriguing questions in Astroparticle Physics concerns 
precisely the origin and nature of the highest energy cosmic rays. 
The existence of events with energy above $10^{20}~$eV has been known since 
the 1960's \cite{EeVents} soon after Volcano Ranch, the first large air 
shower array experiment, started operation. 
Since then they have been slowly but steadily detected by different 
experiments as illustrated in Fig.~\ref{uptonow}. 
The observation of high energy cosmic rays 
has been recently reviewed by Nagano and Watson \cite{WatsonNagano} 
who have shown that there is very good agreement between different 
experiments including the low and high energy regions of the spectrum. 
By now over 17 published events above $10^{20}$~eV \cite{zaspuebla} and 
preliminary new events from HiRes \cite{sokolsky} are enough to 
convince the last skeptics about the non observation of the Greisen-Zatsepin-Kuz'min (GZK) cutoff, expected because of proton 
interactions with cosmic microwave photons \cite{GZK}. 
The data suggest that the spectrum continues 
but little is known about the nature of the arriving particles. 
A large effort is being made in the 2000's to understand these particles.  
A new generation of large aperture experiments has started with 
HiRes already in operation \cite{sokolsky}, the Auger observatory 
in construction and with plans for using new techniques such as 
detection from satellites \cite{EUSO}, with radar  
\cite{gorham} and with radiotelescopes pointing to the 
moon \cite{alvarez} (See Fig.~\ref{uptonow}). 

The Auger project is the last approved large aperture experiment to 
explore the high energy tail of the cosmic ray spectrum, those particles 
with energies exceeding $10^{19}$~eV \cite{AUGER}. The observatory has 
been shown to provide quite large acceptance for the detection of 
inclined showers induced by high energy neutrinos \cite{Capelle}. 
Recent analysis of inclined shower data from the Haverah Park array 
\cite{rate,zasbound} has shown that it is possible to enhance the 
acceptance of air shower arays and to study with them the 
nature of the cosmic ray particles themselves \cite{zasbound}. 
In this article, after briefly addressing shower development, I will 
discuss the main characteristics of the Auger Observatories. The role 
of inclined showers for neutrino detection and composition measurements 
will be stressed, making reference to a model to describe the muon densties 
at ground level induced by inclined cosmic rays. Recent conclusions 
about composition at high energies using vertical and inclined showers 
will be reviewed. 

\section{The Auger Observatories}

As a high energy cosmic ray enters and interacts 
in the atmosphere it gives rise to different generations of secondary 
particles that through succesive interactions constitute the 
extensive atmospheric shower. By the time the shower reaches ground 
level the number of particles in the shower front, mostly photons, electrons 
and positrons can exceed $10^{12}$ for the highest energy cosmic rays. 
As the shower penetrates to further depths the number of photons and  
that of electrons and positrons follow a characteristic behavior not 
too far from a gaussian which reaches a maximum between 1000 and 2000 
meters above sea level for vertical showers above the EeV energy scale. 
Muons arise mainly in the decays of charged pions produced in 
hadronic interactions. 
Unlike electrons, muons do not shower and are practically only subject 
to minimum ionization losses. 
Their depth development increases following the 
pions but hardly decreases after reaching its maximum. 
Typically only muons with energies above the GeV scale reach ground level 
because of decay and energy loss. 
At ground level the photons which are most abundant have an average 
energy of order 1 MeV. Electrons and muons are typically ultrarelativistic 
with respective average energies of order 5 MeV and 1 GeV. 

In the plane transverse to shower axis the shower front has 
a particle density which decreases as the separation 
from shower axis ($r$) increases due to multiple elastic scattering. 
Photons are more numerous than electrons but the lateral distributions 
are similar, decreasing as $r^{-\alpha}$. Although most particles are 
contained within the Moli\`ere radius (of order 100~m),  
for showers above 1~EeV the particle density remains significant 
even when $r$ exceeds one km. 
The muons have a significantly flatter lateral distribution. Although 
they are outnumbered by electrons, the muon density can dominate at large $r$ 
(in the km scale). 
The shower front develops a characteristic curvature depending on the 
position of shower maximum. 
The thickness of the shower front is mostly governed by the different 
delays that the particles accumulate as they deviate from the shower axis. 
Higher energy particles tend to deviate less and thus arrive earlier. 

Extensive air showers have similar distributions whatever the nature of the 
initial particle. The establishment of composition is one of the toughest challenges in the detection of high energy cosmic rays. Simulations reveal 
some differences that can be used for this purpose. One of them is 
the total number of muons in a shower relative to electrons.  
Showers induced by photons mainly cascade into electrons and photons and 
only ocasionally a photon photoproduces mostly pions, channelling part of 
its energy into a hadronic shubshower. As a result 
showers induced by photons typically have of order ten times 
fewer muons than showers induced by protons of the same energy. 
If the cosmic rays are hadrons, the number of muons serves as a discriminator between heavy and light nuclei, the former having a somewhat 
higher muon content. 

\begin{figure}
\center{
\centerline{\hspace{0.2in} \epsfig{file=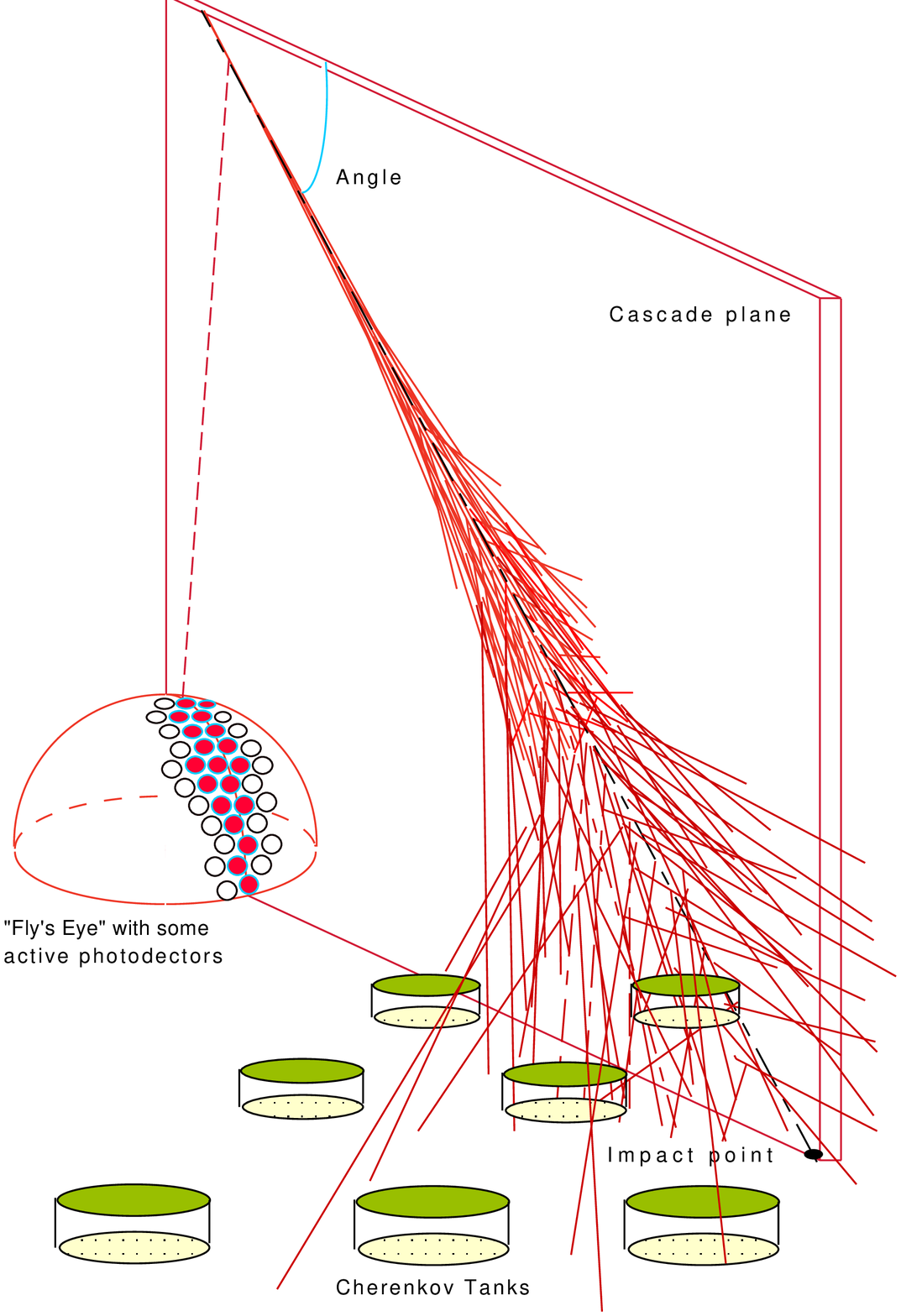,height=3.in,width=2.6in} 
\epsfig{file=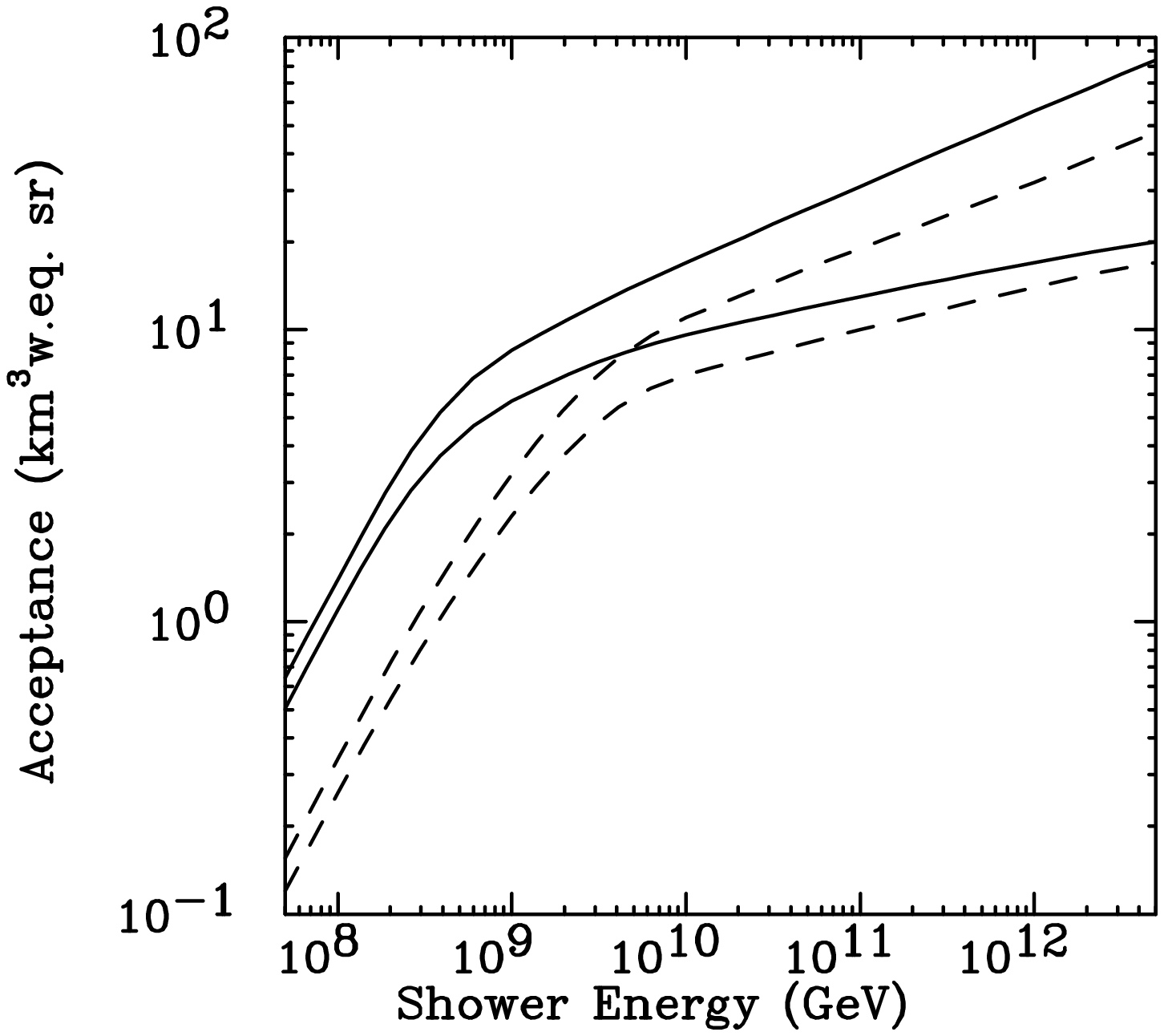,height=2.5in,width=3.2in}} }
\caption{Left: Shower detected by tanks and by a Fluorescent light 
detector or ``eye". Right: The full (dashed) line is the acceptance of the 
Auger Array for detecting the electrons and photons produced 
in hadronic (electromagnetic) showers of $\theta > 60^\circ$. 
Lower curves refer to showers wih axis falling on the array.}
\label{hybrid}
\end{figure}

The Auger project is conceived as two 
3,000 km$^2$ twin observatories in the northern and southern hemispheres, situated at mid latitudes. 
Each observatory is a hybrid experiment combining 
the two only succesful techniques for the study of EeV cosmic 
rays up to now, namely an array of particle detectors and a 
fluorescent detector (see Fig.~\ref{hybrid}). 
The {\sl Engineering Array} is now 
being constructed in {\sl Pampa Amarilla} in Mendoza, Argentina. 
It is a fraction of the Southern Auger array covering only 55~km$^2$ 
and which should be finished during the year 2001. The first 
physics results could be coming very soon. 
The Northern observatory is planned to be sited in Utah, U.S.A. 

The ground array uses cylindrical water 
\v Cerenkov detectors of 10~m$^2$ surface area and 1.2~m of 
height, each instrumented with three photodetectors. 
They are going to be arranged in a hexagonal grid separated 1.5~km 
from each other and extending over a surface area of 3,000 km$^2$. 
Two such tanks have been running in coincidence with the AGASA array.  
When electrons and photons 
reach the tank they are typically absorbed and they give a \v Cerenkov 
light signal which is proportional to the total energy carried by them. 
On the other hand most muons travel through the whole tank and give 
a light signal that is proportional to their track length within the tank. 
The arrival directions of the incident cosmic rays 
is determined from the arrival times of the shower front, typically with 
one degree accuracy. 
Each of the tanks is powered with a solar cell and the data are 
transmitted from the detectors to a central station by conventional 
wireless communication technology. Final triggering will be made at 
the central station.  

The relative 
contributions of each particle species to the tank signals are 
comparable for a proton initiated shower because 
the average energies carried by photons, electrons and muons 
partly compensate the diferences in particle density.  
As we approach shower core both the relative number of muons and the time 
spread of the signal become smaller. At distances of order 1.5~km however  
the signals in the tanks can spread over times of order 2 $\mu$s. At large 
distances to shower axis the relative signal induced by muons becomes 
larger than that of electrons and photons. 
As the signal will be digitized with few nanosecond 
timing, the identification of large \v Cerenkov spikes from the individual 
muons will allow their separation, particularly well away from the shower 
core. 

The particle density 
and the arrival times serve for the determination of the shower energy 
and its direction. 
The ground array has a large aperture, a $100 \%$ duty cycle and a uniform 
right ascension exposure for the study of anisotropy. 
The shower energy determination is usually determined by 
the particle density at a given distance to shower axis. Although this 
distance is chosen to minimize the effects of fluctuations and 
different primaries, the results are unavoidably dependent to some extent 
on both the nature of the arriving particle and the interaction model 
used for shower simulation.  

As the shower develops in the atmosphere, nitrogen 
is excited and emits fluorescence photons in proportion to the number of 
ionizing particles, a few photons per meter of ionizing track length. 
By measuring the light signal and its arrival time from different positions 
along the shower depth development it is possible to detect and study very 
high energy cosmic rays. The fluorescence technique requires mirrors with 
imaging capabilities and covering a sufficient field of view to capture 
the depth development of the shower. The arrival direction 
of the cosmic ray particle is determined by the geometry and timing of 
the arriving light. This method has been very succesful at detecting EeV 
cosmic rays in dark nights. 

The fluorescence detector in the Southern observatory will consist on four 
``eyes", three near the perimeter of the surface array and one roughly in 
the central part, at locations which are slightly elevated with respect 
to the rest of the detector. These eyes are located to 
monitor the atmosphere on top of the ground array.
In the current design each eye is based on Schmidt optics 
and consists of mirror modules each with a 1.7~m 
diaphragm preceeding a 7~m diameter mirror and limiting the  
field of view of each mirror to approximately a 
$30^\circ \times 30^\circ$ fraction of the sky \cite{bluemer}. 
Each eye views $30^\circ$ upwards from the horizon and
combines several of these mirror modules in the azimuthal directions. 
The central eye requires 12 modules to cover $360^\circ$ in azimuth 
and the other 
three eyes to be sited on the perimeter of the ground array only require 
6 or 7 to cover $180^\circ$ or $210^\circ$ in azimuth \cite{dawson}. 
The focal plane of each mirror is instrumented with a camera, an array 
of $20 \times 22$ optical modules, each of viewing aproximately 
$1.5^\circ \times 1.5^\circ$. 

The detection of $10^{20}$~eV showers with the fluorescence technique 
necessarily requires the collection of light which is produced over 
20~km away and thus subject to significant dispersion and attenuation 
in the atmosphere. Moreover the conversion of light to shower signal 
is affected by uncertainties in the geometrical reconstruction of the 
shower direction. Much of the geometrical uncertainty in the reconstruction 
is eliminated if the showers are detected from two or more eyes, that 
is from at least two different locations. This is the {\sl stereo viewing} 
which constitutes one of the important advantages of HiRes \cite{sokolsky}. 
In the Auger observatory many of the 
showers will be viewed in stereo and even by three eyes. 

The fluorescence detector measures the particle production as a function of 
depth into the atmosphere and it is therefore a calorimetric energy determination. The uncertainty in the energy determination is therefore 
reduced with respect to a particle array which is measuring the 
particle content at a 
particular point in shower development and thus is subject to fluctuations 
between different showers. The ability to follow the depth development 
of the shower is also an important advantage because shower maximum can 
be determined directly. Such measurements of depth of maximum are most 
important for the establishment of primary composition. 

The Auger observatory will be the first hybrid detector combining the 
fluorescence technique with a ground array for the detection of EeV 
cosmic rays. The combination of the two techniques is an improtant step 
because besides adding all the features of the two techniques,  
it will serve for cross calibration. The angular resolution of the 
fluorescence technique improves when used in combination with the ground 
array, because much of the geometrical uncertainty in the reconstruction 
of the shower profile is eliminated when the impact point of the shower 
axis is determined by the ground array. The power for composition of using 
both the method of establishing the depth of maximum and that of measuring 
muon content will help to eliminate part of the ambiguity associated to the interdependence between composition and interaction models. 

\section{Inclined Showers and Composition}

The fact that high energy particles exceeding the GZK cutoff 
have been observed allows one to make a strong 
case for the existence of high energy neutrinos. Although the origin 
and nature of these particles is unknown it is difficult to conceive the 
observed flux of any particle species without the existence of a flux of 
high energy neutrinos. 
The detection of inclined showers has been for long 
known to be a possible way to detect very high energy neutrinos 
interacting in the atmosphere \cite{berezinskii}. 
When a neutrino interaction happens deep 
into the atmosphere, the showe can reach its maximum 
very close to ground level in spite of being close to horizontal. 
Such a shower would look much like an ordinary vertical shower with high 
electron and photon content and a front curvature corresponding to 
shower maximum near ground level. 
The Auger observatory will be sensitive to high energy neutrinos. 
Its acceptance for the electromagnetic 
component of deep and inclined showers induced by neutrinos 
exceeds 10~km$^3$sr of water equivalent \cite{Capelle} 
(See Fig.~\ref{hybrid}). 

The original motivation for the study of 
inclined showers induced by cosmic rays was to understand the 
background of cosmic ray signals to neutrino detection, but  
these showers have proved to be of great interest on their own. 
The study of cosmic ray showers by particle arrays 
has been mostly restricted to relatively vertical showers, typically 
resticting zenith angles to less than $45^{\circ}$. 
The particle densities in such such showers keep 
the characteristic circular symmetry allowing an easy reconstruction 
of the event energy by measuring it at a given distance to the shower 
axis. 
The acceptance ${\cal A}$ of a ground array of area $A$ for cosmic rays 
depends on $\theta_{max}$, the maximum zenith angle that the array can 
detect: 
\begin{equation}
{\cal A}= \int A \cos \theta ~ d(\sin \theta)~d \phi = \pi A [1-\cos ^2 \theta _{max}]
\end{equation}
If only zenith angles below $\theta _{max}=45^\circ$ are analysed 
with the Auger observatory, its acceptance would be 4,500 km$^2$sr. 
The acceptance of the observatory will double if the analysis of 
showers can extend to zenith angles less than $90^\circ$. 
It has recently become quite clear that inclined showers produced by 
cosmic rays can be analysed at least with arrays of water \v Cerenkov 
tanks \cite{HSmodel,rate,zasbound}. Moreover the analysis 
of these showers has also shown to have a remarkable potential for 
the study of primary composition \cite{zasbound}. 

Much development has been possible by separately modelling the distortion 
of the muon density patterns in inclined showers under the influence of 
the Earth's magnetic field \cite{HSmodel}. 
The showers can first be studied in the absence of a magnetic field 
where two important facts emerge for inclined showers:  
a) Most of the muons are produced in a well 
defined region of shower development which is quite distant from the ground 
and b) the lateral deviation of a muon is inversely correlated with its energy. 
Most of the characteristics of the muon densities in inclined showers 
are governed by the distance and depth travelled by the muons which is   
of order 4 km for vertical showers, becomes 16 km at 60$^\circ$ 
and continues to rise as the zenith angle rises to exceed 300 km for 
a completely horizontal shower. This distance determines the minimum 
energy needed for a muon to reach ground level and thus fixes the average 
energy of these muons that can be of a few hudred GeV. 

In the absence of a magnetic field we can assume that all muons are 
produced at a given altitude 
$d$ with a fixed transverse momentum $p_\perp$ 
that is uniquely reponsible for the muon deviation from shower axis. 
In the plane transverse to shower axis at ground level (transverse plane) 
the muon deviation, 
$\bar r$, is inversely related to muon momentum $p$. 
The density pattern has full circular symmetry. 
When the magnetic field effects are considered 
the muons deviate a further distance $\delta x$ in the perpendicular 
direction to the magnetic field projected onto the transverse plane 
$\vec B_\perp$, given by:
\begin{equation}
\delta x = \frac {e \vert B_\perp \vert d^2} {2p} = 
\frac{0.15 \vert B_\perp \vert d}{p_{\perp}} \; \bar r = \alpha \; \bar r,
\label{alpha}
\end{equation}
where in the last equation $B_{\perp}$ is to be expressed in Tesla, 
$d$ in m and $p_{\perp}$ in GeV. 

Eq.~\ref{alpha} is telling us that all positive 
(negative) muons that in the absence of a magnetic field would 
fall in a circle of radius $\bar r$ around shower axis, are translated a distance $\delta x$ to the right (left) of the $\vec B_{\perp}$ direction.   
As the muon deviations are small compared to $d$ they can be 
added as vectors in the transverse plane and the muon density pattern is 
a relatively simple transform of the circularly symmetry pattern. 
The dimensionless parameter $\alpha$ measures the relative effect of the 
translation. For small zenith angles $d$ is relatively small and 
$\alpha << 1$ so that the magnetic effects are also small, and results 
into slight elliptical shape of the isodensity curves. 
For high zeniths however $\alpha >1$ the magnetic translation exceeds 
the deviation the muons have due to their $p_{\perp}$ and {\sl shadow} 
regions with no muons appear as confirmed by simulation. 
For an approximate 
$p_{\perp} \sim 200$~MeV and $B_{\perp} = 40~\mu$T this happens when 
$d$ exceeds a distance of order 30~km, that is for zeniths above 
$\sim 70^{\circ}$. 
The muon patterns in the transverse plane can be projected onto the ground 
plane to compare with data as well as standard simulation programs. 
Realistic density patterns are obtained if these ideas are modified 
accounting for the energy distributions of the muons at a given $\bar r$. 
In the simulations it is the average muon energy which is inversely related 
to $\bar r$. 

For each zenith angle the shape of the lateral distribution of the 
muons does not change for showers of energy spaning over four orders 
of magnitude. Different primary particles and interaction models 
also have similar distribution functions in shape. As a result one only 
needs the total number of muons to describe a shower of given zenith. 
This normalization scales with the proton energy $E$ as:
\begin{equation}
N=N_{ref}~E^\beta
\label{Escaling}
\end{equation}
where $\beta$ and $N_{ref}$ are slightly model dependent constants 
\cite{rate}. 


The inclined shower data obtained in the Haverah Park array was 
analysed with the help of the model described above. The Haverah Park 
detector was 
a 12~km$^2$ air shower array using 1.2~m deep water \v Cerenkov 
tanks that was running from 1974 until 1987 in Northern England 
which has been described elsewhere \cite{haverah}. It is the 
array that has been made closest to the ground array of the Auger 
observatory because it also consisted on water \v Cerenkov tanks 
Particular care was taken to account for new corrections to the 
tank signals that arise when horizontal events are detected. 
These include light that falls directly into the photoubes,  
enhanced delta ray signal because muons are more energetic, 
catastrophic energy losses for the muons and  
a signal due to electromagnetic particles from muon decay. 

The event rate as a function of zenith angle 
has been simulated using the modelled muon distributions. 
The qualitative behaviour of the registered rate is well described in the 
simulation and the normalization is also shown to agree with 
data to better than $30\%$ using the measured cosmic ray spectrum for 
vertical incidence, assuming proton primaries and using the QGSM 
model \cite{rate}. 
More impressive are the results of fits of the models for muon densities 
to the observed particle densities sampled by the different detectors on 
an event by event basis. This result demonstrates that the acceptance 
of these detectors can be extended to practically all zenith angles. 
The analysis of the nearly 10,000 events recorded with zenith angles 
above $60^{\circ}$ is a complex process that involves a sequence of time 
fits to get the arrival directions and density fits to obtain the energy 
under the assumption that the primaries are protons. 
The curvature of the shower front is considered and care is taken to 
account for the correlations between the shower energy and the impact point 
of the shower. 
\begin{figure}
\centerline{\epsfig{file=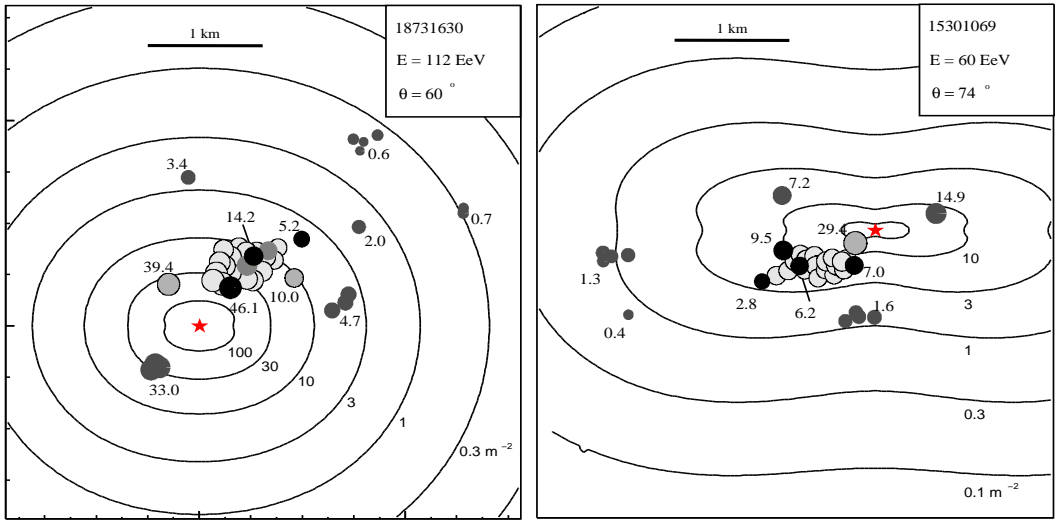,height=2.8in,width=5.6in}}
\caption{Density maps of two events in the plane perpendicular 
to the shower axis. Recorded muon densities are shown as circles 
with radius proportional to the logarithm of the density. The detector
areas are indicated by shading; the area increases from white to black as
1, 2.3, 9, 13, 34 m$^{2}$. The position of the best-fit core is
indicated by a star. Selected densities are also marked. The y-axis 
is aligned with the component of the magnetic field perpendicular to the
shower axis.}
\label{events.fig}
\end{figure}

The analysed data are subject to a set of quality cuts: the shower 
is contained in the detector (distance to core less than 2~km), 
the $\chi ^2$ probability of the event is greater 
than $1 \%$ and the downward error in the reconstructed energy is less than 
$50 \%$. These cuts ensure that the events are correctly reconstructed and 
exclude all events detected above $80^\circ$. Examples of reconstructed 
events compared to predictions are illustrated in Fig.~\ref{events.fig}. 
Two new events with energy exceeding $10^{20}$~eV have been revealed. 
The results have been 
compared to a simulation that reproduces the same fitting 
procedure and cuts using the cosmic ray spectrum deduced from vertical 
air shower measurements in reference \cite{WatsonNagano}.
The agreement between the integral rate above $10^{19}$~eV 
measured and that obtained with simulation is striking when 
the QGSJET model is used for the interactions. 
Sibyll leads to a slight underestimate \cite{zasbound}. 

The universality of the muon lateral distribution function 
is very powerful and once the equivalent proton energy is determined 
for all events, corresponding energies can be easily obtained for 
different assumptions about primary composition. 
In the case the incoming particles are iron nuclei (photons), the 
primary energy can be calculated multiplying the equivalent proton 
energy by a factor which is $\sim 0.7$ (6) for $10^{19}$~eV and varies 
slowly as the primary energy raises. 
As a result when a photon primary spectrum is 
assumed, the simulated rate seriously underestimates the observed data 
by a factor between 10 and 20. A fairly robust bound on the 
photon composition at ultra high energies can be established assuming a
two component proton photon scenario. The photon 
component of the integral spectrum above $10^{19}~$eV (4 $\times 10^{19}$~eV) 
must be less than $41\%$ ($65\%$) at the $95\%$ confidence level. 
Details of the analysis are presented in \cite{zasbound} and will be 
expanded elsewhere. 

\section{Summary}

The Auger detector will soon give an important contribution to the 
observation of the high energy tail of the cosmic ray spectrum. 
Its acceptance is close to 9,000~km$^2$sr when considering the 
inclined air showers. 
Its hybrid character will be of great value to cross calibrate the 
two classes of detectors and for establishing the composition. 
The combined analysis of vertical and 
horizontal showers will also set important limits to composition at high 
energies. The combination of two different techniques for composition 
studies will be of great importance in reducing the uncertainties associated 
to the hadronic interaction models. 

\section*{Acknowledgements}

The author thanks the organizers of such a pleasant conference bringing 
together people from so many different fields, an also thanks D.~Saltzberg 
for helpful comments after carefully reading the manuscript.
This work was supported in part by the European 
Science Foundation (Neutrino Astrophysics Network N. 86), by the CICYT (AEN99-0589-C02-02) and by Xunta de Galicia (PGIDT00PXI20615PR).

\end{document}